# MineProt: modern application for custom protein curation

Yunchi Zhu[1,2,†,*], Chengda Tong[2,†], Zuohan Zhao[2,†], Zuhong Lu[1,2,*]

[1] State Key Laboratory of Bioelectronics, Southeast University, Nanjing, Jiangsu, 210096, China
[2] School of Biological Science and Medical Engineering, Southeast University, Nanjing, Jiangsu, 210096, China

* To whom correspondence should be addressed. Email: huiwenke@seu.edu.cn.
Correspondence may also be addressed to zhlu@seu.edu.cn.

† The authors wish it to be known that, in their opinion, the first 3 authors should be regarded as joint First Authors.

## ABSTRACT

AI systems represented by AlphaFold are rapidly expanding the scale of protein structure modelling data, and the MineProt project provides an effective solution for custom curation of these novel high-throughput data. It enables researchers to build their own protein server in simple steps, run almost out-of-the-box scripts to annotate and curate their proteins, visualize, browse and search their data via a user-friendly online interface, and utilize plugins to extend the functionality of server. It is expected to support researcher productivity and facilitate data sharing in the new era of structural proteomics. MineProt is open-sourced at https://github.com/huiwenke/MineProt.

## INTRODUCTION

AI-based protein structure prediction systems represented by RoseTTAFold [1] and AlphaFold2 [2] have brought the dawn of the high-throughput era of structural proteomics. With accuracy not inferior to traditional methods such as x-ray crystallography and cryo-electron microscopy, they have overwhelming advantages in cost, efficiency and ease of operation [1]. As of July 2022, AlphaFold Protein Structure Database [3] has open access to over 200 million protein structures, a thousand times the scale of PDB's 50-year accumulation. Moreover, thanks to contributions from the open-source community, many optimized versions for AlphaFold have been developed, among which ColabFold [4] is undoubtedly a standout. It significantly reduces the resource requirements for protein folding, enabling more researchers to conduct studies and further enlarging the data scale.

When AI systems are rapidly generating structure predictions, questions arise about how to curate these novel high-throughput data. It is difficult for public database like AlphaFold DB to curate all new data without delays, while it might be hard for most researchers to develop a similar website, where users can search protein information by search engine and analyse their structures by online visualization module. Here we present MineProt, an open-sourced project to help researchers set up a custom protein server with graphical interface in simple steps. An overview of MineProt features is provided below.

## IMPLEMENTATION

MineProt is designed as a server-client architecture. The server is developed by PHP, supporting container deployment including Docker and Podman. It provides a graphical interface for users to curate and analyse their protein structures, which can be grouped into different repositories for better management, and several APIs (application programming interface) for interaction among different modules. Mol* [5] in the style of AlphaFold DB serves as the structure visualization module. It marks the pLDDT of each residue with different colours, and enables users to zoom, rotate, or take a screenshot. Elasticsearch (7.12.1) is employed as the search engine, where protein annotations serve as keywords in its indices. US-align [6] is the key module for structure alignment, supporting PDB structure searching in addition to keyword-searching or BLAST. Besides, MAXIT (11.100) is used for PDB-CIF format converter, which is essential for Mol* visualization.

The client called MineProt toolkit provides several Python3 and Shell scripts for remote management of protein repositories. Interacting with APIs of the server, it can implement a variety of functions such as creating repository, format conversion, protein annotation and data importing.

## WEB INTERFACE

Users will be able to access the MineProt interface's Search Page (**Fig 1 A**) in a web browser via the server IP address, where all their protein repositories will be listed on the left of the page. One or more repositories can be selected for restricting the searching scope and protein information can be retrieved by typing keywords in the search box above powered by Elasticsearch. The results will be demonstrated in the middle of the page, including annotation, structure visualization and links to downloadable file.

Users are able to search for similar structures via the Salign Page (**Video S1**). Searching scope can be restricted by selecting repositories, entering keywords or setting maximum RMSD. Once receiving query structure and parameters from the Salign Page, a background process will be created for alignment to all candidate structures in target repositories. The result page will present Mol* visualization of eligible alignments, where query structure is grey and hit structure is in the style of AlphaFold DB, as well as other standard outputs of US-align.

The MineProt interface is also reinforced by a Browse Page and an Import Page (**Fig S1**). The Browse Page is used to run over protein data in lists, while the Import Page is designed to generate command lines for data importing, especially useful when server administrators are new to MineProt scripting.

## TOOLKIT

The core function of MineProt toolkit can be summarized as transforming results of AI systems to protein repositories compatible with MineProt's modules, as illustrated in **Fig 1 B**. For instance, if a user employs ColabFold to generate large scale structure predictions, it only needs one command to invoke MineProt toolkit to serially unzip result files, convert PDB files to CIF format supporting Mol* visualization, send MSAs (multiple sequence alignments) to UniProt API [7] for homology-based annotations which are transformed into keywords in Elasticsearch indices, and upload all MSAs (.a3m), structures (.pdb, .cif) and model scores (.json) to the target repository. The current version of MineProt toolkit can directly transform raw outputs of AlphaFold and ColabFold into a fully functional website, while supporting any other data compatible with its standard as well.

In addition, the MineProt toolkit also provides plugins for functional extension. For example, given that researchers may expect to search their proteins via BLAST, we have developed a JavaScript plugin linking SequenceServer [8] hits to the MineProt search interface (**Video S2**). It can either be installed in a web browser via TamperMonkey plugin or directly embedded in the SequenceServer application.

## CONCLUSION

It is expected for MineProt to support researcher productivity and facilitate data sharing in the high-throughput era of structural proteomics. Further development of MineProt will continue following the trends in this field. It will support more AI systems, provide more plugins to extend its functionality, update its modules for more efficiency, and become more user-friendly. Joint efforts should be made to transform the high-accuracy protein structure predictions for people's well-being, thus let MineProt be the link among us.

## AVAILABILITY

MineProt is an open-source collaborative initiative available in the GitHub repository (https://github.com/huiwenke/MineProt). Documents are available at https://github.com/huiwenke/MineProt/wiki.

## FUNDING

This research was funded by the National Key Research and Development Project (6307030004).

## ACKNOWLEDGEMENT

We are grateful for Dr. Xiaojun Xia's technical support in the developing & testing environment.

## FIGURES

**Figure 1.** MineProt interface and toolkit. **A.** The Search Page of MineProt interface. **B.** Basic workflow of MineProt toolkit.

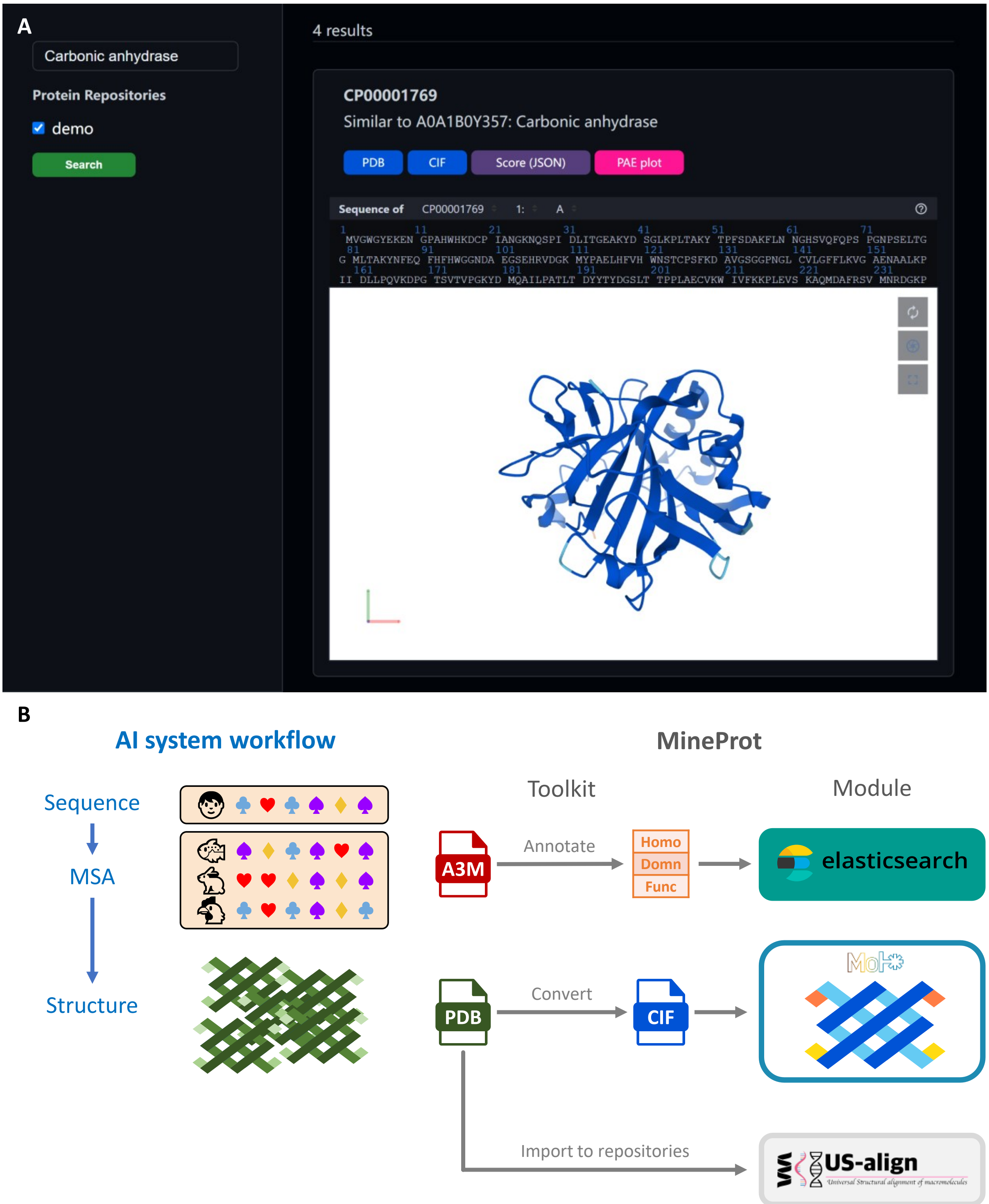

## SUPPLEMENTARY DATA

**Figure S1.** The Browse Page and Import Page of MineProt interface.

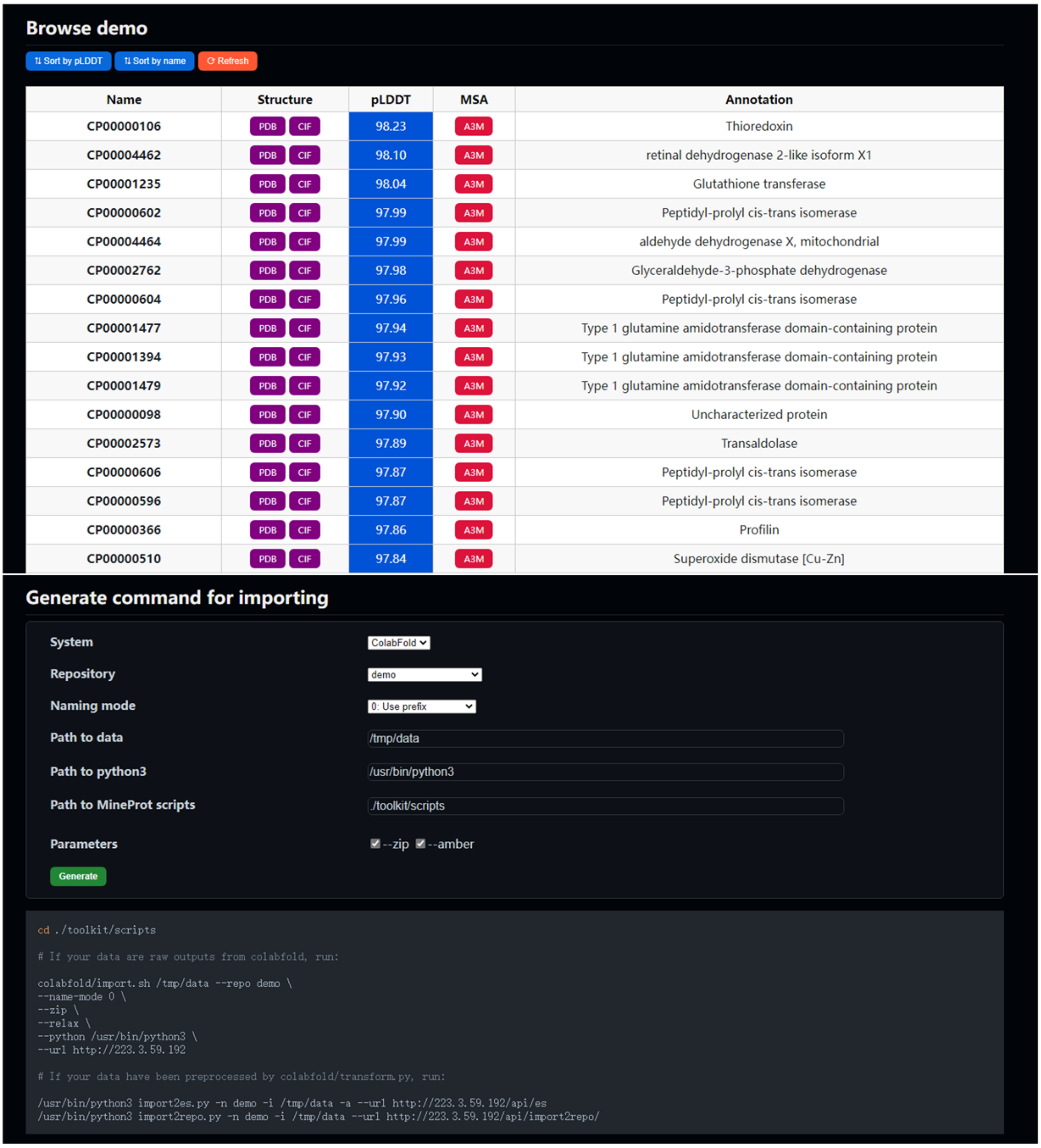

**Video S1.** Demo of MineProt structure alignment service.

https://raw.githubusercontent.com/huiwenke/imgbed/master/mineprot_salign.gif

**Video S2.** Demo of MineProt plugin for SequenceServer.

https://raw.githubusercontent.com/huiwenke/imgbed/master/mineprot_plugin_seqserver.gif